\documentstyle [12pt,aps,amsfonts] {revtex}
\input epsf
\newcommand{\beq}{\begin{equation}}
\newcommand{\eeq}{\end{equation}}
\newcommand{\beqs}{\begin{eqnarray}}
\newcommand{\eeqs}{\end{eqnarray}}

\newcommand{\gsim}{\mathrel{\raisebox{-.6ex}{$\stackrel{\textstyle>}{\sim}$}}}

\begin{document}
\tighten
\draft

\baselineskip 6.0mm

\title{On High-Energy Behavior of Cross Sections in Theories with Large Extra
Dimensions} 

\vspace{8mm}

\author{
Shmuel Nussinov$^{(a,b)}$ \thanks{email: nussinov@post.tau.ac.il} \and
Robert Shrock$^{(b)}$ \thanks{email: robert.shrock@sunysb.edu}}

\vspace{6mm}

\address{(a) \ Sackler Faculty of Science \\
Tel Aviv University \\
Tel Aviv, Israel} 

\address{(b) \ C. N. Yang Institute for Theoretical Physics \\
State University of New York \\
Stony Brook, N. Y. 11794, USA}

\maketitle

\vspace{10mm}

\begin{abstract}

We discuss the high-energy behavior of cross sections in theories with large
extra dimensions and low-scale quantum gravity, addressing two particular
issues: (i) the tension of the D-branes, and (ii) bounds on the cross section
and their relation to approximations in the mode sum over Kaluza-Klein-graviton
exchanges.

\end{abstract}

\vspace{16mm}

\pagestyle{empty}
\newpage

\pagestyle{plain}
\pagenumbering{arabic}
\renewcommand{\thefootnote}{\arabic{footnote}}
\setcounter{footnote}{0}

Theories with large compact dimensions and an associated low scale
characterizing quantum gravity have been the subject of intensive study
recently (a few references are given in \cite{low,comp}).  In Ref. \cite{comp}
we analyzed some of the phenomenological aspects of such theories, including
the high-energy behavior of cross sections for reactions involving graviton
exchange.  Assuming that there are $n$ large compact dimensions of size $\sim
r_n$, and denoting the scale of quantum gravity in the $(4+n)$-dimensional
space as $M_{4+n} \gsim M_s$ (where $M_s$ is the string mass scale in an
underlying string theory), one finds the relation
\beq
M_{Pl}^2 = (r_n)^n (M_{4+n})^{2+n}
\label{mplrel}
\eeq
where $M_{Pl}=G_N^{-1/2} = 1.2 \times 10^{19}$ GeV is the Planck mass, obtained
from the measured Newton constant $G_N$.  Thus, in these theories, the
largeness of $M_{Pl}$ is seen to be a consequence of the largeness of the
compact dimensions, and the underlying short-distance Planck mass, $M_{4+n}$
can actually be much less than $M_{Pl}$.  At large distances $r >> r_n$ (here
for simplicity we assume the same compactification radius for all of the $n$
compact dimensions), the gravitational force $F \propto r^{-2}$, while for $r
<< r_n$, this changes to $F \propto r^{-(2+n)}$. It is a striking fact that
there is a vast extrapolation of 31 orders of magnitude between the smallest
scale of 0.2 mm to which Newton's law has been tested \cite{gravexp} and the
scale that has conventionally been regarded as being characteristic of quantum
gravity, namely the Planck length, $L_{Pl} = 1/M_{Pl} \sim 10^{-33}$ cm,
and it is quite possible that new phenomena could occur in these 31 decades
that would significantly modify the nature of gravity.  It therefore
instructive to explore how drastically one can change the conventional scenario
in which both gauge and gravitational interactions occur in four-dimensional
spacetime up to energies comparable to the Planck mass.  String/brane theory
involves extra dimensions and, especially in Type I constructions, naturally 
lead to models in which open strings end on D$_p$ branes \cite{dbrane}, so
that the standard-model gauge and matter fields propagate on these branes, 
while closed strings (gravitons) propagate in the ``bulk'' between the
branes. 

In these types of models, gravitational scattering is strongly altered by the
presence of a large number of Kaluza-Klein (KK) modes for the graviton.  To an
observer in the usual four-dimensional space, the massless graviton is replaced
by a set of KK modes, of which the lowest is the massless graviton itself, but
the others are massive.  The mass of a KK graviton mode is 
\beq
\mu_{\ell_1,...,\ell_n} = \Bigl ( \sum_{i=1}^n \ell_i^2 \Bigr )^{1/2}
r_n^{-1}
\label{mukk}
\eeq
where the mode numbers are $\ell_j \in {\mathbb Z}$ and the indices $j=1,2,..n$
run over the number of extra compactified dimensions.  The couplings of all of
these KK states to other particles have the same Lorentz structure as the
coupling of the graviton.  These KK-graviton modes have the effect of changing
the strength of gravitational scattering from its naive value $\sim G_N =
M_{Pl}^{-2}$ to a level not much less than regular weak strength, if the
underlying quantum gravity scale $M_s$ is not much greater than the electroweak
scale, $M_{ew} \sim 250$ GeV.  This can be seen because in calculating the rate
for such a scattering process, one multiplies the squared coupling
$M_{Pl}^{-2}$ by a factor incorporating the multiplicity of the various KK
modes that are exchanged.  Since this factor is $\sim (s^{1/2}r_n)^n$, where
$s^{1/2}$ is the center-of-mass energy, when one substitutes the expression for
$r_n$ from eq. (\ref{mplrel}), the factor of $1/M_{Pl}^2$ is exactly cancelled,
and the final product is $s^{n/2}/M_{4+n}^{n+2}$.  Thus, from a
four-dimensional viewpoint, although the KK-gravitons are coupled extremely
weakly, this is compensated by their very large multiplicity, so that their net
effect involves in the denominator a mass scale of order $M_{4+n}$
\cite{trans}.

In Ref. \cite{comp} we carried out estimates of the effects of the exchange of
the KK-graviton modes to $2 \to 2$ gravitational scattering processes.  In the
theories of interest here, as $\sqrt{s}$ becomes comparable to the string
scale, $M_s$, one changes over from a field theory (with effects of D-branes
included) to a fully stringlike picture, so that $M_s$ serves as an upper
cutoff to the low-energy effective field theory in which the calculation is
performed.  Accordingly, one imposes an upper cutoff
\beq
\ell_i < \ell_{max} = M_s r_n
\label{ellmax}
\eeq
on the sums over KK modes, which thus run over the range
\beq
\ell_i = 0, \pm 1, ... , \pm \ell_{max} \quad {\rm for} \quad i=1,...,n
\label{kkrange}
\eeq
The value of $\ell_{max}=M_sr_n$ is very large; for example, for 
$n=2$, for $M_s \sim 30$ TeV, one has
$\ell_{max} = M_s r_2 \sim 4 \times 10^{14}$.

In \cite{comp} we studied several approaches to estimating the high-energy
behavior of the 2-2 gravitational scattering cross section and chose an eikonal
approximation that automatically produced a unitary result for this cross
section.  Our result was (see eq. (2.35) in Ref. \cite{comp})
\beq
\sigma(s)=\frac{4 \pi s}{M_{4+n}^4}
\label{sigma}
\eeq
where $s$ is the center of mass energy squared.  It was emphasized that because
of the universal nature of the gravitational coupling, this applies to any 2-2
scattering process, independent of particle type.  

One application of this result is to ultra-high energy (UHE) neutrinos with
energies ranging from $\sim 10^{15}$ eV to $10^{20}$ eV and beyond.  Neutrinos
in these energy regions can be produced in several ways, including (i) from
active galactic nuclei; (ii) as decay products of the pions produced by the
reaction $p + \gamma_{CMB} \to N + \pi$, where $\gamma_{CMB}$ denotes a photon
in the cosmic microwave background radiation; (iii) via analogous reactions in
which the protons scatter off of the radiation field of a source such as a
gamma-ray burster; and (iv) via decays of Z bosons produced resonantly from
antineutrinos scattering on relic neutrinos \cite{nurev}.  From our estimates,
in conjunction with standard-model calculations of ultra-high energy $\nu p$
and $\bar\nu p$ total scattering cross sections \cite{quigg}, we concluded that
in theories with large compactification radii and low-scale quantum gravity, it
is possible for gravitational scattering to make a non-negligible contribution
to these cross sections.  This could also affect the opacity of the earth to
ultra-high energy neutrinos.  Indeed, from standard-model calculations, it is
known that at energies $E \gsim 10^{15}$ eV, the interaction length for
(anti)neutrinos is smaller than the diameter of the earth, and for $E \gsim
10^{18}$ eV, the earth is opaque to (anti)neutrinos \cite{quigg}, so that the
highest-energy (anti)neutrinos would be expected to arrive at a detector from
the upward hemisphere, while those coming upward and thus traversing more of
the earth are commensurately more highly absorbed enroute. If, indeed, new
gravitational contributions to ultra-high energy $\nu p$ and $\bar\nu p$
scattering are significant, this would be important for currently operating
large detectors like AMANDA and BAIKAL, and the next-generation detectors like
AUGER, NESTOR, ANTARES, ICECUBE, etc. \cite{nurev}.

It is of continuing interest to investigate further what predictions an assumed
underlying string theory of quantum gravity could make concerning models having
extra dimensions whose compactification radii are large.  Unfortunately, at the
present stage of understanding, one cannot derive the compactification process
{\it ab initio} or reliably calculate the compactification radii.  Another
property is the tension of the D$_p$ branes \cite{dbrane} that serve as the
$p$-dimensional spaces, forming $(p+1)$-dimensional world-volumes, in which the
standard-model fields propagate, while gravitons propagate in the
higher-dimensional ``bulk'' extending outside of these D$_p$ branes.  This
tension (more precisely, energy per unit $p$-dimensional spatial volume of the
D-brane) $\tau_p$ is given by \cite{polchinski}
\beq
\tau_p = \frac{1}{g_s (2\pi)^p (\alpha^\prime)^{(p+1)/2}}
\label{taup}
\eeq
where $\alpha^\prime = M_s^{-^2}$ is the Regge slope, which sets the scale of
the string mass $M_s$ and is related to the string tension $T$ according to $T
= 1/(2 \pi \alpha^\prime)$.  Because of the inverse dependence on the string
coupling $g_s$, the D-brane tension, where one can calculate it reliably using
perturbative string theory (i.e., for small $g_s$), satisfies $\tau_p >>
(M_s)^{p+1}$.  More generally, since nonperturbative effects can be important,
one expects that $\tau_p$ is at least of order $(M_s)^{p+1}$.  In turn, this
determines the rigidity of the D-branes; for large tension, recoil effects
accompanying KK-graviton emission are negligibly small.  It follows that
KK-graviton recoil effects are negligible up to the energy $E \sim M_s$ at
which the low energy pointlike effective field theory methods used in
phenomenological studies \cite{low,comp} cease to apply and must be supplanted
by full string-theoretic calculations.  Consequently, in most analyses
\cite{low,comp}, recoil effects are usually dropped. Indeed, even if one takes
a moderately strong string coupling $g_s \sim O(1)$, $\tau_p$ would still be of
order $(M_s)^{p+1}$.  In the exceptional case where one assumes that $\tau_p <<
(M_s)^{p+1}$, the recoil effects would suppress KK-graviton emission and hence
various cross sections involving these \cite{bando,kp}.

Our eikonal approximation used for calculating the KK-graviton 2-2 scattering 
cross section automatically yields a unitary result, and, indeed, this is one
reason that we relied upon this method for our estimate \cite{comp}.  As we
noted in \cite{comp}, the Froissart bound does not apply to KK-graviton
scattering, because this bound assumes that the lightest particle exchanged
in the $t$-channel has a finite mass, but this is not the case for reactions
involving KK-graviton exchange, since the lowest-mass particle is the usual
massless graviton.  Indeed, if one lets $m_{min}$ be the lowest-mass
particle exchanged in the $t$-channel and $s_0$ be a reference value for the
center of mass energy squared, then the Froissart bound is
\cite{froissart,frev} 
\beq
\sigma < \frac{4\pi}{m_{min}^2}\ln^2(s/s_0)
\label{froissart}
\eeq
Thus, clearly, there is no bound if $m_{min}=0$, as is the case here.  

Since the enhanced size of the cross sections involving KK-graviton exchange
comes from the large multiplicity of KK modes, one might inquire what sort of
result one would get by making an approximation in which one excludes the
single massless graviton from the sum over KK modes.  Recall that these modes
are indexed by the $n$ additional KK momenta $k_i = \ell_i/r_n$ where $\ell_i
\in {\mathbb Z}$ and the graviton itself has $\ell_i=0$ for $i=1,...,n$.  In
this case there is a finite, though very small, gap, $m_{min}=1/r_n$.  In
\cite{comp} we concentrated on the case $n=2$, so let us take this again here
for definiteness; in this case, $r_2=M_{Pl}/M_6^2$.  The Froissart bound
then allows (neglecting the logs) $\sigma=4\pi /m_{min}^2= 4 \pi r_2^2$. This
is exactly the value of our linearly rising cross section (\ref{sigma}) at
$s=M_{Pl}^2$.  In this case there are just four KK-graviton exchanges that are
characterized by the minimum nonzero mass, namely, for $n=2$, $k_\lambda = (\pm
1/r_2,0), \ (0,\pm 1/r_2)$.  The strength of the partial wave amplitude
$s/M_{Pl}^2$ of each exchange is order unity at this value $s=M_{Pl}^2$,
thereby allowing for the saturation (up to logs) of the Froissart bound.

Consider next some lower energy $s=M_{Pl}^2/N^2$. The relevant KK modes are the
$O(N^2)$ modes satisfying the condition $k_1^2+k_2^2 < (N/r_2)^2$, i.e. the
modes that have squared masses smaller than $(N/r_n)^2$. In this case the
Froissart bound allows cross sections $\sigma=4 \pi r_2^2/N^2$, again in
agreement with our eq. (\ref{sigma}).  Furthermore the reduction of each
individual exchange amplitude by the factor $s/M_{Pl}^2 = 1/N^2$ is compensated
by the multiplicity of the relevant $\sim N^2$ KK-graviton exchanges, so that
again we have a total resultant amplitude of order unity and hence saturation
of the Froissart bound \cite{trans}.  Of course for the true cross section, one
cannot exclude the contribution of the massless graviton, which, as noted,
means that the conventional Froissart bound does not apply to this cross
section.  In passing, we note that in Ref. \cite{kp} an attempt to analyze the
high-energy behavior due to KK-graviton was made, and seemed to give a somewhat
weaker growth of the cross section than in (\ref{sigma}).  We have traced the
reason for this; instead of the $t$-dependence $1/(t-\mu_{\ell_1,...\ell_n}^2)$
appropriate for the exchange of the KK-graviton mode with extra-dimensional
momenta $(\ell_1/r_n,..., \ell_n/r_n)$, Ref. \cite{kp} used an {\it ad hoc}
exponential form.  As the derivation of the Froissart bound clearly shows, it
is the $t$-dependence (equivalently, the dependence on the impact parameter,
$b$) that matters and not the initial $s^\nu$ behavior of the Born amplitude
that, even for $\nu \ge 2$, is removed by unitarization in systems with finite
range interactions.

\vspace{4mm} 

S. N. would like to thank the Israeli Academy for Fundamental Research for a
grant.  The research of R. S. was partially supported by the NSF grant
PHY-97-22101.  S. N. thanks the Yang Institute for Theoretical Physics at SUNY
Stony Brook and the physics department, UCLA, and R. S. thanks Tel Aviv
University for hospitality during visits when part of this paper was written.

\vfill
\eject


\begin{thebibliography}{99}
\bibitem{dbrane}{Here a D$_p$-brane is an extended object with $p$ spatial
dimensions, on which a string ends \cite{polchinski}.  For the usual
superstring $X(\sigma,\tau)^\mu$ in $d=10$ spacetime, the boundary conditions
for a single D$_p$-brane can be taken to be $\partial_\sigma X^\mu=0$ (Neumann)
for $\mu=0,1,...,p$ and $X^\mu=0$ (Dirichlet) for $\mu=p+1,...,9$.}
\bibitem{trans}{Note that the enhancement of the cross section all the way up
to the level $\sim 10^{-32}$ cm$^2$ at $E_\nu = 10^{21}$ eV of significance to
ultra high energy neutrino interactions does not require extremely small
momentum transfers and can involve KK-graviton momentum transfers $|t| \sim$ 
TeV$^2$.}
\end{thebibliography}
\end{document}